\documentclass[prb,twocolumn,aps,showpacs,epsf,dvips]{revtex4}   %
\usepackage{feynmp}
\usepackage{amssymb}
\usepackage{epsfig}
\usepackage{graphicx}
\usepackage{subfigure}
\usepackage{color}
\begin{document}

\title{Interaction and excitonic insulating transition in graphene}
\author{Guo-Zhu Liu, Wei Li, and Geng Cheng \\
{\small {\it Department of Modern Physics, University of Science
and Technology of China, Hefei, Anhui, 230026, P.R. China }}}

\begin{abstract}
The strong long-range Coulomb interaction between massless Dirac
fermions in graphene can drive a semimetal-insulator transition. We
show that this transition is strongly suppressed when the Coulomb
interaction is screened by such effects as thermal fluctuation,
doping, disorder, and finite volume. It is completely suppressed
once the screening factor $\mu$ is beyond a threshold $\mu_{c}$ even
for infinitely strong coupling. However, such transition is still
possible if there is an additional strong contact four-fermion
interaction. The differences between screened and contact
interactions are also discussed.
\end{abstract}


\maketitle


The low-energy elementary excitations in undoped graphene are
massless Dirac fermions. Their spectral and transport properties are
quite unusual and have attracted intense investigations in the past
several years \cite{Geim, CastroNeto}. For a clean undoped graphene,
the density of states (DOS) $N(\omega)$ vanishes linearly near the
Dirac point. As a result, the Coulomb interaction between massless
Dirac fermions is essentially unscreened, in sharp contrast to the
electron system with parabolic dispersion. The unscreened,
long-range Coulomb interaction was shown to be responsible for many
anomalous behaviors of graphene \cite{Gonzalez94, Khveshchenko2001,
Gusynin02, Khveshchenko2006, Herbut06, Vafek07, Son07, Aleiner,
Sheehy}.

At the strong coupling regime, the long-range Coulomb interaction
can open a finite mass gap for the Dirac fermion, which then drives
a phase transition from the semimetal state to an insulator state.
This transition is realized by forming stable particle-hole pairs
and usually named as excitonic semimetal-insulator (SM-IN)
transition \cite{Khveshchenko2001, Gusynin02}. Recently, this kind
of phase transition has been studied by nonperturbative
Dyson-Schwinger (DS) equation approach \cite{Khveshchenko2001,
Gusynin02, Khveshchenko2006}, renormalization group method
\cite{Son07}, and lattice simulations \cite{Hands08, Drut09}. The
SM-IN transition was found in graphene for strong Coulomb coupling
and small fermion flavor \cite{Khveshchenko2001, Gusynin02}. The
effects of finite temperature and external magnetic field were also
considered \cite{Gusynin02}.

Although being of remarkable interests, the predicted SM-IN
transition (in zero magnetic field) has not yet been unambiguously
observed in experiments. In this paper, we discuss the effects that
can potentially prevent the appearance of this SM-IN transition.
First of all, it should be emphasized that such transition can take
place only for strong, poorly screened Coulomb interaction.
Generically, there are two critical parameters: critical
dimensionless coupling strength $\lambda_{c}$ and critical fermion
flavor $N_{c}$. SM-SI transition is possible only when $N < N_{c}$
and $\lambda > \lambda_{c}$. Once the long-range Coulomb interaction
is screened by some physical effects, there will be an effective
screening factor $\mu$, which is expected to increase $\lambda_{c}$
and reduce $N_{c}$. This can be understood by noting the important
fact that SM-SI transition realized by forming fermion-antifermion
pairs is a genuine low-energy phenomenon. From the experience in
QED$_{3}$, the long-range nature of gauge interaction plays the
crucial role in generating the dynamical mass gap for initially
massless Dirac fermions \cite{Liu03}. A finite gauge boson mass
rapidly reduces the critical fermion flavor to below the physical
value $2$ \cite{Liu03}. In the present case, there is a similar
suppressing effect once the long-range Coulomb interaction is
screened for some reason. The opening of excitonic gap requires that
the Coulomb interaction is sufficiently strong at low-momentum
region. However, the screening factor $\mu$ suppresses the
contribution from small momenta significantly. Obviously, this kind
of pairing instability is markedly different from the conventional
BCS-type pairing formation, which is caused by arbitrary weak
attractive force between electrons.

In realistic graphene samples, the critical behavior of SM-IN
transition can be influenced by the following reasons: thermal
fluctuation; doping; disorder; finite sample volume. Each of them
can generate an effective screening factor $\mu$, which could be
regarded as an effective photon mass. We study their effects on
critical strength $\lambda_{c}$ and critical flavor $N_{c}$ by
solving the corresponding gap equation, and show that a growing
$\mu$ significantly increases $\lambda_{c}$ and reduces $N_{c}$,
both at zero and finite temperatures. When $\mu$ is beyond some
threshold $\mu_{c}$, the excitonic transition is completely
prohibited, leaving semi-metal as the stable ground state.
Frequently, some of these effects coexist, leading to further
suppression of excitonic transition. However, even when $\mu >
\mu_{c}$, we found that the excitonic transition can still take
place if there is an additional strong contact quartic interaction.
We also briefly discuss the interesting differences between the
screened Coulomb and contact quartic interactions.

The total Hamiltonian of massless Dirac fermion $H = H_{0} + H_{C}$
is given by
\begin{eqnarray}
&& H_{0} = v_{F}\sum_{\sigma=1}^{N}\int_{\mathbf{r}}
\bar{\psi}_{\sigma}(\mathbf{r})i\mathbf{\gamma}\cdot\mathbf{\nabla}
\psi_{\sigma}(\mathbf{r}), \nonumber \\
&& H_{\mathrm{C}} = \frac{1}{4\pi}\sum_{\sigma,\sigma^{\prime}}^{N}
\int_{\mathbf{r},\mathbf{r}^{\prime}}
\bar{\psi}_{\sigma}(\mathbf{r})\gamma_{0}\psi_{\sigma}(\mathbf{r})
\frac{e^{2}}{|\mathbf{r}-\mathbf{r}^{\prime}|}
\bar{\psi}_{\sigma^{\prime}}(\mathbf{r}^{\prime})\gamma_{0}
\psi_{\sigma^{\prime}}(\mathbf{r}^{\prime}). \nonumber
\end{eqnarray}
Here, we adopt four-component spinor field $\psi$ to describe the
massless Dirac fermion since there is no chiral symmetry in the
two-component representation. The conjugate spinor field is defined
as $\bar{\psi} = \psi^{\dagger} \gamma_{0}$. The $4\times 4$
$\gamma$-matrices satisfy the standard Clifford algebra. Although
the physical fermion flavor is actually $N = 2$, in the following we
consider a large $N$ in order to perform $1/N$ expansion. The total
Hamiltonian preserves a continuous U(2N) chiral symmetry $\psi
\rightarrow e^{i\theta \gamma_{5}}\psi$, which will be dynamically
broken if a nonzero fermion mass gap is generated.

The free propagator of massless Dirac fermion is $G_{0}(k_{0},
\mathbf{k}) = (\gamma_{0}k_{0} - v_{F}\mathbf{\gamma}\cdot
\mathbf{k})^{-1}$. The Coulomb interaction modifies it to the
complete propagator
\begin{eqnarray}
G(k_{0},\mathbf{k}) = \frac{1}{\gamma_{0}k_{0}A_{1}(k) -
v_{F}\mathbf{\gamma}\cdot \mathbf{k}A_{2}(k) - m},
\end{eqnarray}
where $m(k)$ denotes the dynamical fermion mass and $A_{1,2}$ the
wave function renormalization functions. To the leading order in
$1/N$ expansion, the DS integral equation is
\begin{eqnarray}
G^{-1}(p) &=& G_{0}^{-1}(p) + \int \frac{d^{3}k}{(2\pi)^{3}}
\gamma_{0}G(k)\gamma_{0} V(p-k),
\end{eqnarray}
where the vertex function has already been approximated by the bare
matrix $\gamma_{0}$. The nontrivial solution $m(p)$ of this equation
signals the opening of an excitonic gap.

In the DS gap equation, $V(q)$ is the Coulomb interaction function.
The bare, unscreened Coulomb interaction has the form $V_{0}(q) =
\frac{g_C^{2}}{2|\mathbf{q}|}$ in the momentum space. For an
interacting electron gas, the collective density fluctuations screen
the bare Coulomb interaction $V_{0}(q)$ to $V^{-1}(q) =
V_{0}^{-1}(q) - \pi(q)$. For ordinary non-relativistic electron gas,
the static polarization function $\pi(q_{0}=0)$ is just the
zero-energy DOS, $N(0)$, which is known to be finite. The parameter
$N(0)$ defines the inverse Thomas-Fermi screening length. The case
for undoped clean graphene is quite different because of the linear
dispersion of Dirac fermions. The leading contribution to
polarization function is given by $\pi_{0}(q) = -
\frac{N}{8}\frac{\mathbf{q}^{2}}{\sqrt{q_{0}^{2} +
v_F^{2}|\mathbf{q}|^{2}}}$. It vanishes linearly as $\mathbf{q}
\rightarrow 0$ in the static limit $q_{0}=0$, so the long-range
Coulomb interaction is unscreened.

\begin{figure}[t]
  \centering
  \subfigure{
    \label{} 
    \includegraphics[width=2.2in]{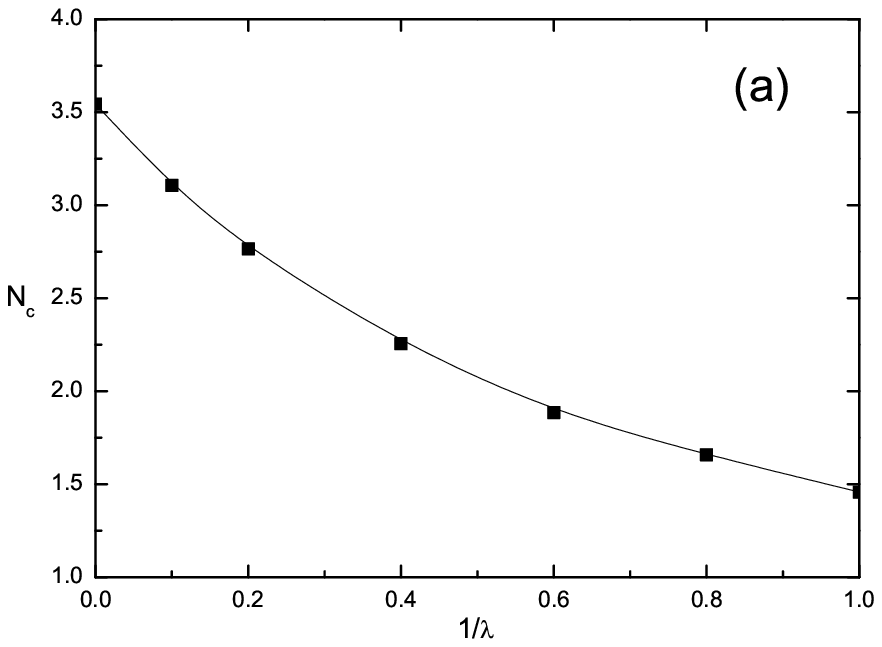}}
  \subfigure{
    \label{} 
    \includegraphics[width=2.2in]{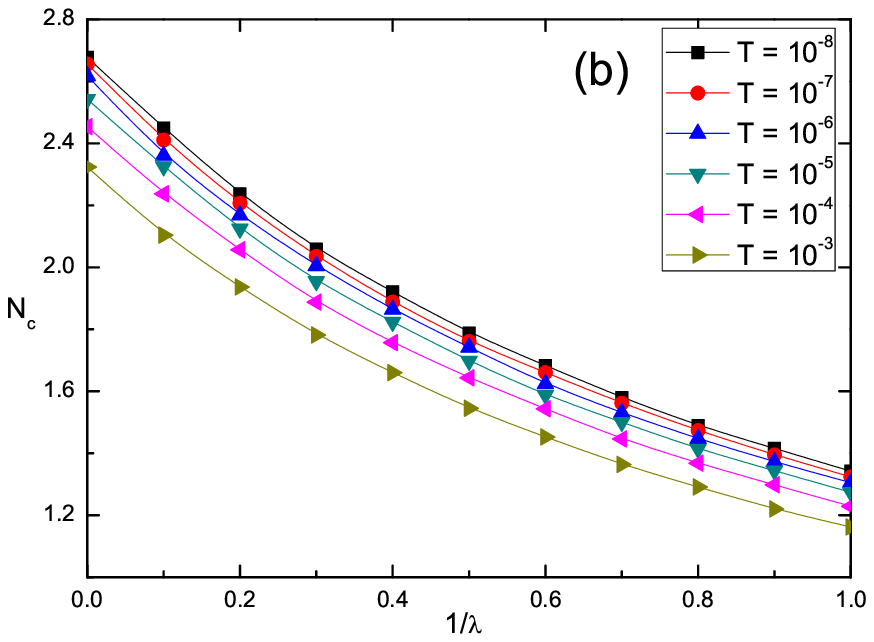}}
 \caption{(a) Relationship between $N_{c}$ and $\lambda$ at zero temperature;
 (b) Relationship between $N_{c}$ and $\lambda$ at different temperatures $T$. Both are for unscreened Coulomb
 interaction.}
  \label{fig:fig1} 
\end{figure}

Under the approximations described above, the gap equation can be
written as
\begin{eqnarray}
\label{eq:gap} m(p^{2}) = \frac{1}{N}\int\frac{dk_0}{2\pi}
\frac{d^{2}\mathbf{k}}{(2\pi)^{2}} \frac{m(k^{2})}{k_{0}^{2} +
|\mathbf{k}|^{2} + m^{2}(k^{2})}V(p-k),
\end{eqnarray}
with interaction function
\begin{eqnarray}
V(q) = \frac{1}{\frac{|\mathbf{q}|}{8\lambda} + \frac{1}{8}
\frac{|\mathbf{q}|^{2}}{\sqrt{q_{0}^{2} + |\mathbf{q}|^{2}}}}.
\end{eqnarray}
Here, $A_{1,2} = 1$ is assumed and the rescaling
$v_{F}\mathbf{k}\rightarrow \mathbf{k}$, $v_{F}
\mathbf{\Lambda}\rightarrow \mathbf{\Lambda}$ is made (such
rescaling will be made throughout the whole paper). The present
problem contains two parameters: fermion flavor $N$ and
dimensionless Coulomb coupling defined as $\lambda =
g_{C}^{2}N/16v_{F}$, where $g_{C} =e^2/\epsilon_{0}$. The
ultraviolet cutoff $\Lambda$ is taken to be of order $10$eV which is
determined by $\sim a^{-1}$ with lattice constant $a = 2.46
\textrm\AA$. Not that no instantaneous approximation for the
polarization function is made at present. We solve the nonlinear gap
equation using bifurcation theory and parameter embedding method
\cite{Liu03, Cheng} for a number of fixed values of $\lambda$. The
fermion flavor $N$ serves as the embedded parameter in seeking the
bifurcation point. The results are shown in Fig.~\ref{fig:fig1}(a).
It is easy to see that the critical flavor $N_{c}$ is an increasing
function of $\lambda$. For $\lambda \rightarrow \infty$, $N_{c}
\approx 3.52$; for $\lambda = 2$, $N_{c} \approx 2$.

\begin{figure}[t]
  \centering
   \subfigure{
    \label{fig:subfig:a} 
    \includegraphics[width=2.2in]{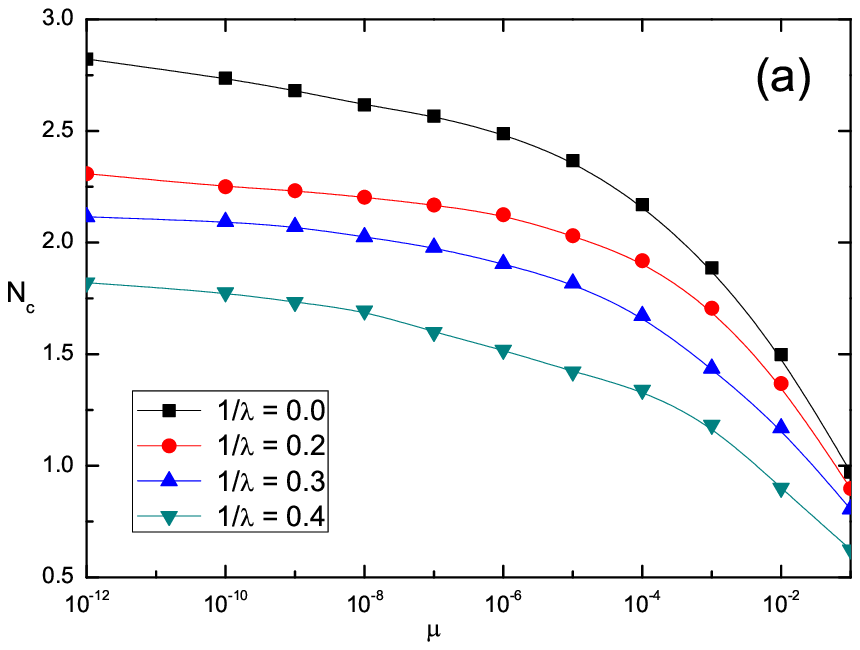}}
  \subfigure{
    \label{fig:subfig:b} 
    \includegraphics[width=2.2in]{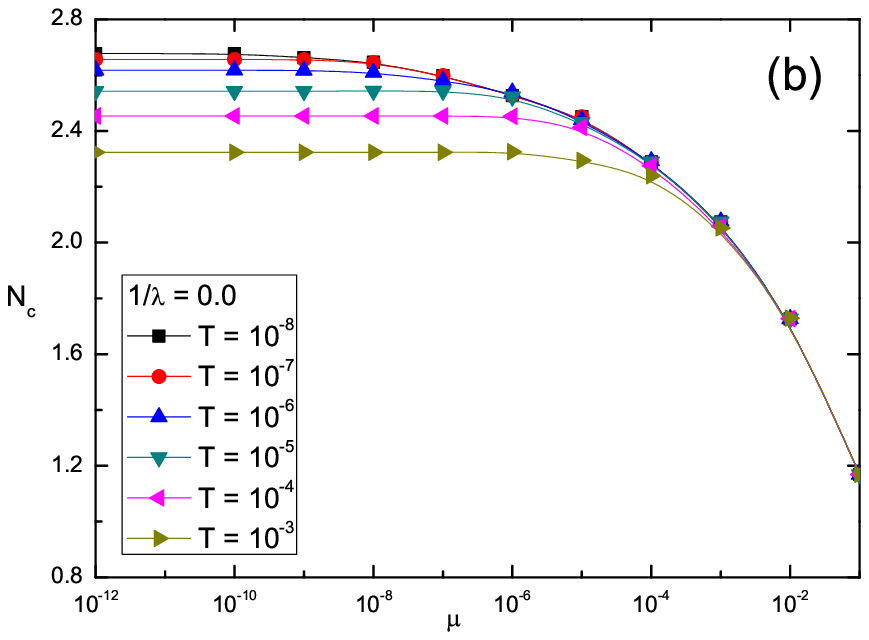}}
\caption{(a) Dependence of $N_{c}$ on $\mu$ for different $\lambda$
at zero temperature; (b) Dependence of $N_{c}$ on $\mu$ for
different $T$ at $\lambda \rightarrow \infty$.}
  \label{fig:fig2} 
\end{figure}

The above results are valid only for the unscreened Coulomb
interaction at zero temperature. In realistic systems, the
long-range interaction could be screened by several physical
effects, such as thermal fluctuation, doping, disorder, and finite
volume. If the polarization function $\pi(q_{0},\mathbf{q})$ takes a
finite value due to some mechanism in the $q_{0}=0$ and
$\mathbf{q}\rightarrow 0$ limit, then the long-range Coulomb
interaction becomes short-ranged and $\pi(0,0)$ defines the
screening factor. Before computing $\pi(q_{0},\mathbf{q})$ by taking
each screening effect into account, we now phenomenologically
introduce a single parameter $\mu$ (in unit of eV) to model the
screened interaction function
\begin{equation}
V(q) = \frac{1}{\frac{|\mathbf{q}|}{8\lambda} + \frac{1}{8}
\frac{|\mathbf{q}|^{2}}{\sqrt{q_{0}^{2} + |\mathbf{q}|^{2}}} + \mu}.
\end{equation}
The advantage of this parameter is that it explicitly measures the
suppressing effect on the critical behavior due to all possible
screening mechanisms. If we regard this function as the effective
interaction strength, then the influence of $\mu$ becomes clear: it
eliminates the contribution of small momenta to the gap equation
Eq.~(\ref{eq:gap}). But remember that the excitonic gap generation
is primarily determined by the contribution from this region, so it
is expected that a large $\mu$ will destroy SM-IN transition. After
solving the gap equation, we found that a growing $\mu$ leads to
increase of critical strength $\lambda_{c}$ and to decrease of
critical flavor $N_{c}$ (see Fig.~\ref{fig:fig2}(a)). Beyond some
critical value $\mu_{c}$, the SM-IN transition is completely
prevented, even when the dimensionless strength $\lambda \rightarrow
\infty$.

The possible screening mechanisms will be discussed in order. First
of all, the thermal fluctuation will surely restore the chiral
symmetry even it is broken by the ground state. At finite
temperatures, the Matsubara fermion propagator is
\begin{eqnarray}
\mathcal{G}(i\omega_{n},\mathbf{k}) = \frac{1}{i\omega_{n}\gamma_{0}
- v_{F}\mathbf{\gamma}\cdot \mathbf{k} - m},
\end{eqnarray}
where $\omega_{n} = (2n+1)\pi T$ is the fermion frequency. Here, in
order to carry out the frequency summation appearing in the gap
equation, we utilize the instantaneous approximation
\cite{Khveshchenko2001, Gusynin02}. Under this approximation, the
polarization function can be approximated \cite{Aitchinson} by
\begin{eqnarray}
\label{eq:finitet_pi} \pi(0,\mathbf{q}) =
\frac{N}{8v_{F}^{2}}\left(v_{F}\mathbf{q} + cT \exp
\left(-\frac{v_F\mathbf{q}}{cT}\right)\right),
\end{eqnarray}
with constant $c=16\ln2/\pi$. At the limit $\mathbf{q} \rightarrow
0$, the polarization is $\sim T$, corresponding to the thermal
screening factor $\mu$. Since other screening effects can coexist
with thermal fluctuations at finite temperatures, we still introduce
the parameter $\mu$ and write the gap equation as
\begin{eqnarray}
\label{eq:finitet_gap}
 m(\mathbf{p},T) &=& \frac{1}{N}\int
\frac{d^{2}\mathbf{k}}{8\pi^{2}}
\frac{m(\mathbf{k},T)}{\sqrt{\mathbf{k}^{2}+
m^{2}(\mathbf{k},T)}}V(\mathbf{p}-\mathbf{k},T) \nonumber \\
&& \times \tanh\frac{\sqrt{\mathbf{k}^{2} +
m^{2}(\mathbf{k},T)}}{2T}.
\end{eqnarray}
with the interaction function
\begin{equation}
\label{eq:finitet_v}
 V(\mathbf{q},T) =
\frac{1}{\frac{\mathbf{|q|}}{8\lambda}+\frac{1}{8} \big(\mathbf{|q|}
+ cT e^{-\frac{\mathbf{|q|}}{cT}}\big) + \mu}.
\end{equation}
The results at finite temperatures are rather complex since now we
have four parameters, $N$, $\lambda$, $T$, $\mu$, each of which has
a critical value. Their relationships are shown in
Fig.~\ref{fig:fig1}(b) without screening effects ($\mu = 0$) and in
 Fig.~\ref{fig:fig2}(b) with screening (the temperature is in unit of
$\mathrm{eV}$). In  Fig.~\ref{fig:fig2}(b), the Coulomb coupling
parameter is fixed at $\lambda \rightarrow \infty$, and the results
for other values of $\lambda$ are not shown since they are
qualitatively similar. The results tell us that the thermal
suppression is more important than screening effect when $\mu$ has
small values ($< 10^{-5}$), but the screening effect eventually
becomes much more important than thermal effect for larger values of
$\mu$.

\begin{figure}[t]
  \centering
  \subfigure{
    \label{fig:subfig:a} 
    \includegraphics[width=2.2in]{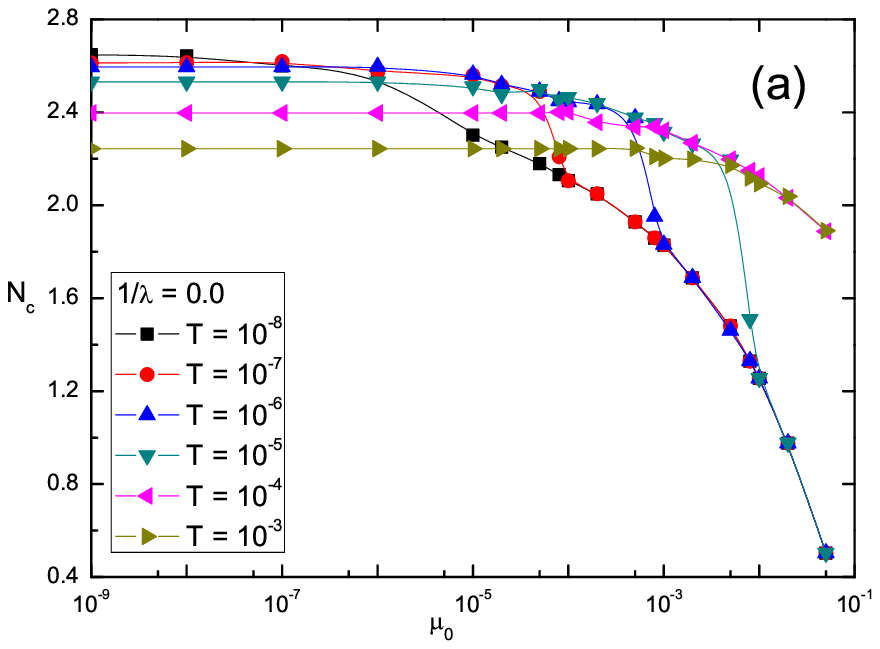}}
  \subfigure{
    \label{fig:subfig:b} 
    \includegraphics[width=2.2in]{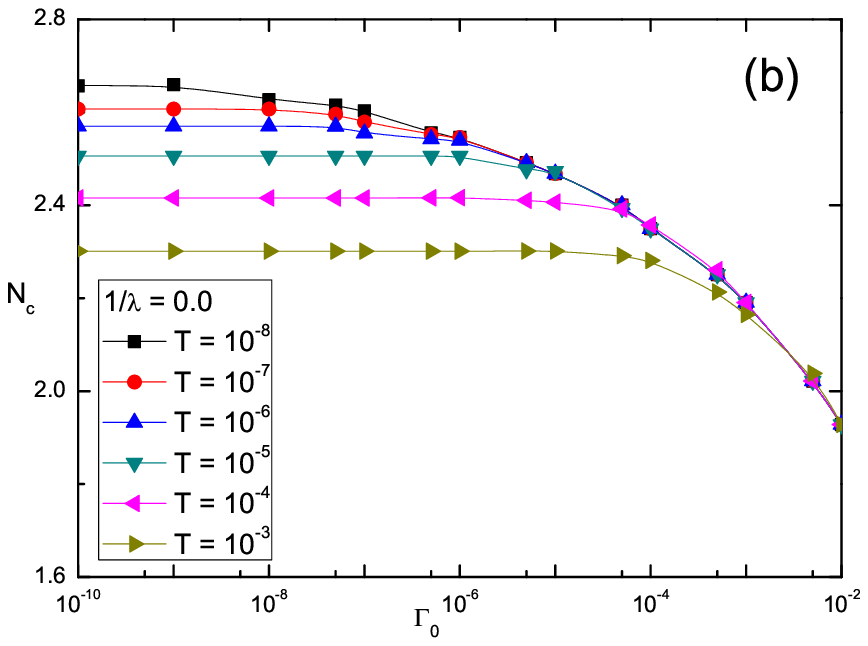}}
 \caption{(a) Dependence of $N_{c}$
on $\mu_{0}$ for different values of $T$ at $\lambda \rightarrow
\infty$; (b) Dependence of $N_{c}$ on $\Gamma_{0}$ for different
values of $T$ at $\lambda \rightarrow \infty$.}
  \label{fig:fig3} 
\end{figure}

The second potential mechanism that can prevent gap generation is
doping. The Coulomb interaction between Dirac fermions is unscreened
only when the graphene is undoped. When the graphene is slightly
doped, the finite carrier density then serves as an effective
screening factor $\mu$. Then the excitonic gap is expected to open
only at or very close to the Dirac point. The critical carrier
density has been discussed previously in \cite{Gusynin02}. Recently,
the same screening effect was emphasized in the study of exciton
condensate in bilayer graphene \cite{Efetov}. At finite chemical
potential $\mu_{0}$, the fermion propagator becomes
\begin{eqnarray}
\mathcal{G}(i\omega_{n},\mathbf{k},\mu_{0}) = \frac{1}{(i\omega_{n}
- \mu_{0})\gamma_{0} - v_{F}\mathbf{\gamma}\cdot \mathbf{k} - m}.
\end{eqnarray}
Using this propagator, the polarization function can be calculated
with the result
\begin{eqnarray}
\pi(0,\mathbf{q},\mu_{0}) &=& \frac{2NT}{v_{F}^{2}}\int_0^1
dx\Big[\ln(2\cosh\frac{\sqrt{x(1-x)\mathbf{q}^{2}} + \mu_{0}}{T})
\nonumber \\
&& + \ln (2\cosh\frac{\sqrt{x(1-x)\mathbf{q}^{2}} -
\mu_{0}}{T})\Big],
\end{eqnarray}
in the zero frequency limit. As $\mathbf{q} \rightarrow 0$,
$\pi(0,0,\mu_{0}) = \frac{2N}{v_{F}^{2}}\mu_{0}$, which defines the
screening factor $\mu$. After performing the frequency summation,
the gap equation has the form
\begin{eqnarray}
\label{eq:mu_gap}
m(\mathbf{p},T) &=& \frac{1}{N}\int
\frac{d^{2}\mathbf{k}}{8\pi^{2}}
\frac{m(\mathbf{k},T)}{\sqrt{\mathbf{k}^{2} +
m^{2}(\mathbf{k},T)}}V(0,\mathbf{p}-\mathbf{k},\mu_{0})
\nonumber \\
&& \times \Big[\frac{1}{e^{\frac{\mu_{0}-\sqrt{\mathbf{k}^{2}+
m^{2}}}{T}}+1} - \frac{1}{e^{\frac{\mu_{0}+\sqrt{\mathbf{k}^{2}+
m^{2}}}{T}}+1}\Big], \nonumber
\end{eqnarray}
with the interaction function being
\begin{equation}
V(0,\mathbf{q},\mu_{0}) = \frac{1}{\frac{\mathbf{|q|}}{8\lambda} +
\frac{1}{N}\pi(0,\mathbf{q},\mu_{0})}.
\end{equation}
Note the chemical potential  $\mu_{0}$ appears in two places: the
occupation number and the polarization function. To see the dominant
effect of $\mu_{0}$, we first solved the full gap equation and show
the results in Fig.~\ref{fig:fig3}(a) (also at $\lambda \rightarrow
\infty$ for comparison). The dependence of $N_{c}$ on $T$ and
$\mu_{0}$ qualitatively resembles that in Fig.~\ref{fig:fig2}(b),
but visibly exhibits different quantitative behavior: the
suppressing effect from doping is more prominent at low $T$ than at
higher $T$. Despite the details, a large doping makes the excitonic
insulating state impossible. Then we solved the gap equation by
ignoring the $\mu_{0}$-dependence of occupation number and found
that the results are nearly the same as Fig.~\ref{fig:fig3}(a)
(therefore not shown). It seems that the screening effect induced by
doping plays the dominant role in suppressing the gap generation.

Next, we consider the influence of disorders, which are unavoidable
in graphene samples. The disorders can be crudely classified as
random mass, random chemical potential, and random vector potential,
\emph{etc}, and have been extensively treated using various field
theoretic techniques \cite{Fradkin, Lee93, Ludwig94, Nersesyan95,
Mirlin08}. The low-energy DOS was found to be sensitive to the
symmetry of disorders \cite{Ludwig94, Nersesyan95, Mirlin08}. For
instance, random vector potential leads the DOS to vanish
algebraically upon approaching the Fermi surface with exponent
depending on symmetry \cite{Ludwig94, Mirlin08}. For this kind of
disorder, there is essentially no screening effect and the Coulomb
interaction remains long-ranged, provided that the Altshuler-Aronov
type correction to low-energy DOS is not included. For random mass
potential, the zero-energy DOS can have finite value, as a result of
dynamical discrete symmetry breaking \cite{Fradkin, Nersesyan95}. In
the case of weak disorders, the impurity scattering can be treated
within the conventional self-consistent Born approximation, which
reveals that the zero-energy DOS acquires a finite value of the form
\cite{Lee93}, $N(0) = \frac{N}{\pi^{2}v_{F}^{2}}\Gamma_{0}
\ln\frac{\Lambda}{\Gamma_{0}}$, with a constant scattering rate
$\Gamma_{0}$. The finite $N(0)$ screens the long-range Coulomb
interaction. Within the Matsubara formalism, such screening effect
can be elaborated by including the scattering rate $\Gamma_{0}$ into
the polarization function.

To study the role of weak disorders, we first write the effective
Dirac fermion propagator
\begin{eqnarray}
\mathcal{G}(i\omega_{n},\mathbf{k},\Gamma_{0}) =
\frac{1}{(i\omega_{n} + i\Gamma_{0}\mathrm{sgn}\omega_{n})\gamma_{0}
- v_{F}\mathbf{\gamma}\cdot \mathbf{k} - m},
\end{eqnarray}
which contains the scattering rate $\Gamma_{0}$. Due to the sign
dependence of scattering rate, the gap equation and polarization
function becomes rather complicated. After frequency summation
within the instantaneous approximation, the gap equation is found to
be
\begin{eqnarray}
\label{eq:gamma_gap} m(\mathbf{p},T) &=& \frac{1}{N}\int \frac{d^{2}
\mathbf{k}}{4\pi^{2}} \frac{m(\mathbf{k},T)}{\sqrt{\mathbf{k}^{2} +
m^{2}(\mathbf{k},T)}}V(0,\mathbf{p}-\mathbf{k},\Gamma_0) \nonumber \\
&& \times \frac{1}{\pi}\mathrm{Im}\Big[\psi\Big(\frac{1}{2} +
\frac{\Gamma_{0}}{2\pi T} + i\frac{\sqrt{\mathbf{k}^{2}+
m^{2}}}{2\pi T}\Big)\Big],
\end{eqnarray}
where $\psi(x)$ is the digamma function. At the clean limit,
$\Gamma_{0}=0$, the imaginary part of digamma function can be
simplified as $\mathrm{Im}[\psi(\frac{1}{2} +
i\frac{\sqrt{\mathbf{k}^{2}+ m^{2}}}{2\pi T})] =
\frac{\pi}{2}\tanh\frac{\sqrt{\mathbf{k}^{2}+ m^{2}}}{2T}$ which is
the same as that appearing in gap equation
Eq.~(\ref{eq:finitet_gap}). As in the case of chemical potential,
the screening effect caused by disorder scattering can be directly
seen by calculating the vacuum polarization function
$\pi(\omega_{n},\mathbf{q},\Gamma_{0})$ and then taking the
$\omega_{n}=0, \mathbf{q}\rightarrow 0$ limit. However, even in the
instantaneous approximation, it is not easy to obtain the complete
form of $\pi(0,\mathbf{q},\Gamma_{0})$. When the scattering rate
$\Gamma_{0}$ is larger than the thermal scale $\sim T$, $\Gamma_{0}
> 2\pi T$, we found that the polarization function can be well
approximated by the following expression(as detailed in Appendix)
\begin{equation}
\label{eq:pi_gamma} \pi(0,\mathbf{q},\Gamma_{0})
\approx\frac{N}{8}(\mathbf{q} +
c^{\prime}\Gamma_{0}\exp(-\frac{\mathbf{q}}{c^{\prime}\Gamma_{0}})),
\end{equation}
with constant $c^{\prime} = \frac{8 10^{\ln2}\ln2}{\pi^{2}}$. At the
limit $\mathbf{q}=0$, it takes a finite value
\begin{equation}
\pi(0,0,\Gamma_{0}) = \frac{10^{\ln2}\ln2}{\pi^{2}}N\Gamma_0,
\end{equation}
which is proportional to the scattering rate $\Gamma_{0}$ and
defines the screening factor. Comparing the polarization
Eq.~(\ref{eq:pi_gamma}) with Eq. (7), formally the scattering rate
$\Gamma_{0}$ plays the role of an effective temperature $T$. Now the
interaction function in gap equation Eq.~(\ref{eq:gamma_gap})
becomes
\begin{equation}
V(0,\mathbf{q},\Gamma_{0}) = \frac{1}{\frac{\mathbf{|q|}}{8\lambda}
+ \frac{1}{N}\pi(0,\mathbf{q},\Gamma_{0})}.
\end{equation}
On the other hand, in the case of small $\Gamma_{0}$ the
polarization function $\pi(0,\mathbf{q},\Gamma_{0})$ should be
replaced by Eq. (7). After solving the full gap equation
Eq.~(\ref{eq:gamma_gap}), we present the dependence of $N_{c}$ on
scattering rate $\Gamma_{0}$ for different values of $T$ in
Fig.~\ref{fig:fig3}(b). In order to see the effects of screening on
gap generation, we also solved the gap equation when $\Gamma_{0}$
appears only in the interaction function
$V(0,\mathbf{q},\Gamma_{0})$. The quantitative difference between
the results in these two cases is negligible. The results in
Fig.~\ref{fig:fig3}(b) show that there is a competition between the
suppressing effects of thermal fluctuation and disorder scattering.
At low temperature $T$, the scattering rate $\Gamma_{0}$ dominates;
while for small $\Gamma_{0}$, the thermal effect dominates.
Obviously, a large $\Gamma_{0}$ suppresses the possibility of gap
generation rapidly. Further, we solved the gap equations
Eq.~(\ref{eq:finitet_gap}) and Eq.~(\ref{eq:finitet_v}) with the
screening factor simply set to be $\mu = N(0) =
\frac{N}{\pi^{2}v_{F}^{2}}\Gamma_{0} \ln\frac{\Lambda}{\Gamma_{0}}$
and found that the results are quantitatively similar to
Fig.~\ref{fig:fig3}(b).

One might argue that the low-energy fermionic excitations are all
suppressed once a fermion mass gap opens, so the DOS vanishes at
energy scale below the gap and there is no screening effect.
However, for fermion of mass $m$, the zero-energy DOS was found
\cite{GusyninEPJ} to be $N(0) = \frac{2}{\pi^{2}v_{F}^{2}}
\Gamma_{0} \ln\frac{\Lambda}{\sqrt{\Gamma_{0}^{2}+m^{2}}}$. In
principle, we might include a gap $m$ into the polarization function
$\pi(0,\mathbf{q},\Gamma_{0},m)$ and then study the gap equation.
Since the critical behavior of SM-IN transition is studied by
linearizing the nonlinear gap equation, the mass can be safely set
to zero, $m \rightarrow 0$, near the bifurcation point.

Finally, we discuss effect of finite sample volume (area in two
dimensions). For a graphene plane of finite spatial extent, the
particle momenta becomes discrete and the momenta transferred in the
process of interaction can not be arbitrary small. If we still work
in the continuum field theoretic formalism, this effect can be
equivalently represented by imposing an infrared cutoff $\kappa$,
given by the inverse sample size $L^{-1}$. Its effects on $N_{c}$ is
nearly the same as  Fig.~\ref{fig:fig2}(a) at $T = 0$ and
Fig.~\ref{fig:fig2}(b) at finite $T$ with $\mu$ replaced by
$\kappa$, and hence are not shown explicitly. The results imply that
the sample of large spatial extent is more favorable to undergo the
SM-IN transition \cite{Gusynin03}.

Besides the above four effects, any other mechanism that can screen
the long-range Coulomb interaction will also unavoidably lower the
possibility of gap generation. If more than one screening effects
coexist in reality, the suppression of SM-IN transition becomes much
more significant, as shown in Fig.~\ref{fig:fig2}. In light of these
results, we conclude that the excitonic insulating state can most
probably be observed in undoped, clean graphene of large area near
absolutely zero temperature.

Once the long-range Coulomb interaction is screened, one interesting
question is whether it can be equivalently replaced by a short-range
or even a contact (on-site) repulsive interaction \cite{CastroNeto}.
This question can also be asked in another way: is the long-range
nature or the strong coupling nature of Coulomb interaction more
important in driving the SM-IN transition? If the answer is the
latter, then the long-range interaction can well be replaced by a
short-range or contact one. According to the above results, it seems
that the long-range, rather than strong coupling, nature plays the
dominant role. As shown in Fig.~\ref{fig:fig2}(a), even in the very
strong coupling limit $\lambda \rightarrow \infty$, the critical
flavor $N_{c}$ is already less than the physical flavor $2$ when the
screening factor $\mu \sim 10^{-3}$. For moderately strong coupling
$\lambda = 2.5$, the excitonic insulating behavior becomes
impossible even if the screening factor is only as small as $\mu
\sim 10^{-12}$.

\begin{figure}[h]
\begin{center}
\includegraphics[width=2.2in]{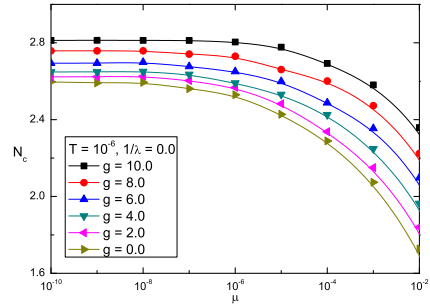}
\end{center}
\caption{Dependence of $N_{c}$ on $\mu$ for different $g$.}
\label{fig:fig4}
\end{figure}

In order to test the role of contact interaction and see its
difference from the screened Coulomb interaction, we add one quartic
interacting term to the Hamiltonian. There are several choices for
the four-fermion coupling term, classified by the gamma matrices
used to define the action \cite{Gross, Rosenstein}. For simplicity,
we consider only one of them, i.e.,
\begin{eqnarray}
\frac{G}{N}\sum_{\sigma}^{N} \int_{\mathbf{r}}
(\bar{\psi}_{\sigma}(\mathbf{r}) \psi_{\sigma}(\mathbf{r}))^{2}.
\end{eqnarray}
To the lowest order, this contact interaction contributes the
following term
\begin{eqnarray}
\frac{g}{N\Lambda}\int \frac{d^{2}\mathbf{k}}{8\pi^{2}}
\frac{m(\mathbf{k},\beta)\tanh\frac{\sqrt{\mathbf{k}^{2}+
m^{2}(\mathbf{k},\beta)}}{2T}}{\sqrt{\mathbf{k}^{2}+
m^{2}(\mathbf{k},\beta)}},
\end{eqnarray}
to the gap equation, where the dimensionless coupling is
$g=NG\Lambda/v_F$ and the scaling $v_F\mathbf{k}\rightarrow
\mathbf{k}$ is made as before. The whole gap equation is solved with
results shown in Fig.~\ref{fig:fig4} at $T = 10^{-6}
\mathrm{eV}$($\sim 10\mathrm{mK}$). The contact four-fermion
interaction has opposite effect on the critical flavor $N_{c}$ as
compared with the screening factor $\mu$: while the latter rapidly
suppresses $N_{c}$, the former is very efficient in promoting the
system towards the excitonic insulating phase (note there is no
Goldstone boson in the insulating phase since the total Hamiltonian
preserves discrete chiral symmetry $\psi \rightarrow
\gamma_{5}\psi$). Thus we see that the contact four-fermion
interaction is actually different from the screened Coulomb
interaction. For a relatively large screening factor $\mu$, the
latter is unable to generate excitonic gap even in the $\lambda
\rightarrow \infty$ limit, while the former can generate such gap
when its coupling is larger than some critical value $g > g_{c}$.
The reason for this difference can be seen from the gap equation:
for screened Coulomb interaction, $\mathbf{q}$ appearing in the
denominator suppresses the contribution from large momenta, while
$\mu$ in the denominator suppresses the contribution from small
momenta; on the contrary, for contact fermion interaction, the
coupling $g$ is constant in the whole momenta region without any
suppressing effect. In conclusion, the SM-IN transition is still
possible if there is additional strong contact fermion interaction,
even when the screened Coulomb interaction itself can not open the
gap.

We end with a brief discussion on the validity of the gap equation
used in this paper. In a rigorous treatment, the excitonic gap
generation should be studied by solving the self-consistent
equations of fermion self-energy function, wave function
renormalization, Coulomb interaction, and vertex function. In
practice, a number of approximations must be utilized. Here we kept
only the Fock diagram for the fermion self-energy and omit all
higher order corrections of the $1/N$ expansion
\cite{Khveshchenko2001, Gusynin02}. The results should be
qualitatively reliable for large $N$. To verify the conclusions
obtained in the leading order, it would be necessary to include
these corrections (such as wave function renormalization, vertex
function correction, \emph{etc.}) since they might change the
quantities of critical parameters $\lambda_{c}$ and $\mu_{c}$
considerably for the physical flavor $N=2$. However, this is beyond
the scope of the present work. Another question concerns the
important effect of velocity renormalization on the excitonic gap
instability \cite{Son07, Aleiner}. This effect has been addressed
recently by incorporating the momentum-dependent fermion velocity
into the gap equation \cite{Khveshchenko09}. It was found that the
velocity renormalization does not dramatically affect the excitonic
instability \cite{Khveshchenko09}.

G.Z.L. thanks D. V. Khveshchenko, T. Tu and I. L. Aleiner for
helpful communications. This work was supported by the NSF of China
under Grant No. 10674122.


\appendix

\section{CALCULATION OF $\pi(0,\mathbf{q},\beta,\Gamma_{0})$}

In this appendix, we present the derivation of the polarization
function at finite impurity scattering rate and finite temperature.
The fermion contribution to the vacuum polarization is given by
\begin{eqnarray}
\pi(\omega_{m},\mathbf{q},\beta) = -\frac{N}{\beta}
\sum_{n=-\infty}^{\infty}\int\frac{d^{2}\mathbf{k}}{(2\pi)^{2}}
\frac{\mathrm{Tr}[\gamma_{0}k\!\!\!/\gamma_{0}(q\!\!\!/+k\!\!\!/)]}
{k^2(q+k)^2}.
\end{eqnarray}
Here $q_{0} \equiv i\omega_{m} = \frac{2m\pi}{\beta}$ and $k_{0}
\equiv i\omega_{n} = \frac{(2n+1)\pi}{\beta}$, and a new momentum
variable is defined by $l = k + xq$ with $l_{0} =
i\omega_{m}+i\omega_{n}$.

Within the instantaneous approximation $\omega_{m}=0$, the
polarization function reduces to
\begin{eqnarray}
\pi(0,\mathbf{q},\beta) = \frac{4N}{\beta} \int_{0}^{1}dx
\int\frac{d^{2}\mathbf{l}}{(2\pi)^{2}} \left[S_{1} -
2\mathbf{l}^{2}S_{2}\right],
\end{eqnarray}
where $S_{i=1,2}$ is given by
\begin{equation}
S_{i} = \sum_{n=-\infty}^{\infty}\frac{1}{\left[l_{0}^{2} +
\mathbf{l}^{2} + x(1-x)\mathbf{q}^{2}\right]^{i}}.
\end{equation}
In the presence of impurity scattering rate $\Gamma_{0}$, the
variable $l_{0}$ should be replaced by
\begin{equation}
l_{0} = \frac{2\pi}{\beta}(n + \frac{1}{2} +
\frac{\beta}{2\pi}\Gamma_{0}\mathrm{sgn}\omega_{n}).
\end{equation}
Using the notation in Ref. \cite{Dorey}, we define a new variable $Y
= \frac{\beta}{2\pi}\sqrt{\mathbf{l}^{2} + x(1-x)\mathbf{q}^{2}}$.
Using the identity
\begin{eqnarray}
S(X,Y) &=& \sum_{n=0}^{\infty}\frac{1}{(n+X)^{2}+Y^{2}} \nonumber \\
&=& \frac{1}{2Yi}\left[\psi(X+iY)-\psi(X-iY)\right],
\end{eqnarray}
the function $S_{1}$ now becomes
\begin{eqnarray}
S_{1} &=& \frac{\beta^{2}}{4\pi^{2}}\left[S(\frac{1}{2} +
\frac{\beta}{2\pi}\Gamma_{0},Y) +
S(1-\frac{1}{2}+\frac{\beta}{2\pi}\Gamma_{0},Y)\right], \nonumber \\
&=& \frac{\beta^{2}}{2\pi^{2}Y}\mathrm{Im}
[\psi(\frac{1}{2}+X^{\prime}+iY)],
\end{eqnarray}
from which the function $S_{2}$ is given by $S_{2} =
-\frac{\beta^{2}}{8\pi^{2}Y}\frac{\partial S_{1}}{\partial Y}$.
Define $t = \frac{2\pi}{\beta}Y$, and $t^{2} \equiv [\mathbf{l}^{2}
+ C_{\mathbf{l}}^{2}] =
[\mathbf{l}^{2}+(\sqrt{x(1-x)\mathbf{q}^{2}})^2]$, then the
polarization function $\pi(0,\mathbf{q},\beta,\Gamma_{0})$ is
written as the following integral
\begin{eqnarray}
&& \frac{2N}{\pi^2} \int_{0}^{1}dx \int_{C_{\mathbf{l}}}^{\infty} dt
\Big[\frac{C_{\mathbf{l}}^{2}}{ t^2}\mathrm{Im}
[\psi(\frac{1}{2}+\frac{\beta}{2\pi}\Gamma_{0} +
i\frac{\beta}{2\pi}t)] \nonumber \\
&& + \frac{t^{2} - C_{\mathbf{l}}^{2}}{t}\frac{\partial}{\partial t}
\mathrm{Im}[\psi(\frac{1}{2}+\frac{\beta}{2\pi}\Gamma_{0} +
i\frac{\beta}{2\pi}t)]\Big].
\end{eqnarray}
It is hard to compute this integral analytically. For relatively
large scattering rate $\Gamma_0$, we found that the $\psi$ function
can be approximated by the analytic expression
\begin{equation}
\psi(\frac{1}{2} + \frac{\beta}{2\pi}\Gamma_{0} +
i\frac{\beta}{2\pi}t) \approx \frac{\pi}{2}\tanh\frac{\pi
t}{10^{\ln2}\Gamma_0}
\end{equation}
for $\frac{\beta}{2\pi}\Gamma_{0} > 1$ (with error $1\%$ for
$\frac{\beta}{2\pi}\Gamma_{0} \gg 1$ and averaging error $5\%$ for
$\frac{\beta}{2\pi}\Gamma_{0} \approx 1$). Then the integration over
variable $t$ can be carried out with the result
\begin{eqnarray}
\pi(0,\mathbf{q},\Gamma_{0}) \approx
\frac{10^{\ln2}N\Gamma_{0}}{\pi^{2}}\int_{0}^{1}dx\,
\ln\big[2\cosh\frac{\pi\sqrt{x(1-x)\mathbf{q}^{2}}}{10^{\ln2}\Gamma_{0}}\big].
\nonumber \\
\end{eqnarray}
It has the similar form as Eq.~(\ref{eq:finitet_pi}) with
$\Gamma_{0}$ playing the role of an effective "temperature", thus
the polarization function can now be approximated by
\begin{equation}
\pi(0,\mathbf{q},\Gamma_{0}) \approx\frac{N}{8}(\mathbf{q} +
c^{\prime}\Gamma_{0}\exp(-\frac{\mathbf{q}}{c^{\prime}\Gamma_{0}})),
\end{equation}
where $c^{\prime} = \frac{8\ln2\,10^{\ln2}}{\pi^2}$. At the clean
limit, $\Gamma_{0}\ll \pi/\beta$, we have
\begin{equation}
\psi(\frac{1}{2} + \frac{\beta}{2\pi}\Gamma_{0} +
i\frac{\beta}{2\pi}t) \approx \frac{\pi}{2}\tanh(\frac{\beta t}{2}).
\end{equation}
In this case, the polarization function is still approximated by
Eq.~(\ref{eq:finitet_pi}).



\begin{thebibliography}{99}

\bibitem{Geim}
A. K. Geim and K. S. Novoselov, Nat. Mater. {\bf 6}, 183 (2007).

\bibitem{CastroNeto}
A. H. Castro Neto, F. Guinea, N. M. R. Peres, K. S. Novoselov, and
A. K. Geim, Rev. Mod. Phys. {\bf 81}, 109 (2009).

\bibitem{Gonzalez94}
J. Gonzalez, F. Guinea, and M. A. H. Vozmediano, Nucl. Phys. B {\bf
424}, 595 (1994).

\bibitem{Khveshchenko2001}
D. V. Khveshchenko, Phys. Rev. Lett. {\bf 87}, 246802 (2001); D. V.
Khveshchenko and H. Leal, Nucl. Phys. B {\bf 687}, 323 (2004).

\bibitem{Gusynin02}
E. V. Gorbar, V. P. Gusynin, V. A. Miransky, and I. A. Shovkovy,
Phys. Rev. B {\bf 66}, 045108 (2002).

\bibitem{Khveshchenko2006}
D. V. Khveshchenko and W. F. Shively, Phys. Rev. B {\bf 73}, 115104
(2006).

\bibitem{Herbut06}
I. F. Herbut, Phys. Rev. Lett. {\bf 97}, 146401 (2006); O. Vafek and
M. J. Case, Phys. Rev. B {\bf 77}, 033410 (2008).

\bibitem{Vafek07}
O. Vafek, Phys. Rev. Lett. {\bf 98}, 216401 (2007).

\bibitem{Son07}
D. T. Son, Phys. Rev. B {\bf 75}, 235423 (2007).

\bibitem{Aleiner}
I. L. Aleiner, D. E. Kharzeev, and A. M. Tsvelik, Phys. Rev. B {\bf
76}, 195415 (2007).

\bibitem{Sheehy}
D. E. Sheehy and J. Schmalian, Phys. Rev. Lett. {\bf 99}, 226803
(2007).

\bibitem{Hands08}
S. J. Hands and C. G. Strouthos, Phys. Rev. B {\bf 78}, 165423
(2008).

\bibitem{Drut09}
J. E. Drut and T. A. Lahde, Phys. Rev. Lett. {\bf 102}, 026802
(2009); Phys. Rev. B {\bf 79}, 165425 (2009).

\bibitem{Liu03}
G.-Z. Liu and G. Cheng, Phys. Rev. D {\bf 67}, 065010 (2003).

\bibitem{Cheng}
G. Cheng and T. K. Kuo, J. Math. Phys. {\bf 35}, 6270 (1994); {\bf
35}, 6693 (1994).

\bibitem{Aitchinson}
I. J. R. Aitchinson, N. Dorey, M. Klein-Kreisler, and N. E.
Mavromatos, Phys. Lett. B {\bf 294}, 91 (1992).

\bibitem{Efetov}
M. Yu. Kharitonov and K. Efetov, Phys. Rev. B {\bf 78}, 241401(R)
(2008). See also a comment by R. Bistritzer, H. Min, J. J. Su, and
A. H. MacDonald, arXiv:0810.0331.

\bibitem{Fradkin}
M. P. A. Fisher and E. Fradkin, Nucl. Phys. B {\bf 241}, 457 (1985);
E. Fradkin, Phys. Rev. B {\bf 33}, 3263 (1986).

\bibitem{Lee93}
P. A. Lee, Phys. Rev. Lett. {\bf 71}, 1887 (1993); A. Durst and P.
A. Lee, Phys. Rev. B {\bf 62}, 1270 (2000).

\bibitem{Ludwig94}
A. W. W. Ludwig, M. P. A. Fisher, R. Shankar, and G. Grinstein,
Phys. Rev. B {\bf 50}, 7526 (1994).

\bibitem{Nersesyan95}
A. A. Nersesyan, A. M. Tsvelik, and F. Wenger, Nucl. Phys. B {\bf
438}, 561 (1995).

\bibitem{Mirlin08}
A. Altland, B. D. Simons, and M. R. Zirnbauer, Phys. Rep. {\bf 359},
283 (2002); F. Evers and A. D. Mirlin, Rev. Mod. Phys. {\bf 80},
1355 (2008).

\bibitem{GusyninEPJ}
V. P. Gusynin and V. A. Miransky, Eur. Phys. J. B {\bf 37}, 363
(2004).

\bibitem{Gusynin03}
The similar effect was studied in QED$_{3}$ by V. P. Gusynin and M.
Reenders, Phys. Rev. D {\bf 68}, 025017 (2003).

\bibitem{Gross}
D. J. Gross and A. Neveu, Phys. Rev. D {\bf 10}, 3235 (1974).

\bibitem{Rosenstein}
B. Rosenstein, B. J. Warr, and S. H. Park, Phys. Rep. {\bf 205}, 59
(1991).

\bibitem{Khveshchenko09}
D. V. Khveshchenko, J. Phys: Condens. Matter {\bf 21}, 075303
(2009).

\bibitem{Dorey}
N. Dorey and N. E. Mavromatos, Nucl. Phys. B {\bf 386}, 614 (1992).


\end{thebibliography}
\end{document}